\documentclass[twoside]{article}

\usepackage{graphicx}
\usepackage{amsmath}
\usepackage{multicol}

\makeatletter

\newenvironment{figurehere}
  {\def\@captype{figure}}
  {}
\makeatother

\begin{document}
\setlength{\textheight}{8.78truein} 

\thispagestyle{empty}

\markboth{A. Vidybida}
{Testing of Information Condensation}

\setcounter{page}{1}

\vspace*{1.05truein}

\begin{center}
{\large\bf  Testing of information condensation in a model\\[0.04truein]
reverberating spiking neural network\footnote{Accepted in the International Journal of Neural Systems, http://www.worldscinet.com/ijns/ijns.shtml}}
\end{center}
\vspace*{0.04truein}
\vspace*{0.45truein}
\centerline{A. K. Vidybida}
\vspace*{0.0215truein}
\centerline{\it Department of Synergetics,
Bogolyubov Institute for Theoretical Physics}
\centerline{\it 	      Metrologichna str., 14-B, 03680 Kyiv, Ukraine}
\baselineskip=11pt
\centerline{\it E-mail: vidybida@bitp.kiev.ua}

\vspace*{0.29truein}
\abstract{Information about external world is delivered to 
the brain in the form of structured in time
spike trains. During further processing in higher areas, 
information is subjected to a certain
condensation process, which results in formation
of abstract conceptual images of external
world, apparently, represented as certain uniform 
spiking activity partially independent on
the input spike trains details. Possible physical mechanism of condensation
at the level of individual neuron was discussed recently. 
In a reverberating spiking neural network, due to this mechanism 
the dynamics should settle down to the same
uniform/periodic activity in response to a set of various inputs. Since the same
periodic activity may correspond to different input spike trains, we interpret
this as possible candidate for information condensation mechanism in a network.
Our purpose is to test this possibility in a network 
model
consisting of five fully connected  neurons,
particularly, the influence of geometric size of the network, on its ability to
condense information.
Dynamics of 20 spiking neural networks of different geometric sizes are modelled
by means of computer simulation. Each network was propelled into reverberating
dynamics by applying various initial input spike trains. We run the dynamics until it 
becomes periodic. The Shannon's formula is used to calculate the amount of information
 in any input spike train and in any periodic state found. As a result,  we obtain explicit estimate of  the degree of information condensation in the networks, and conclude that it depends strongly on the net's geometric size.}{}
\vspace*{10pt}

\begin{multicols}{2}
\section{Introduction}
\noindent
The ability of a biological object to obtain enough information about external world is essential for the object's survival. The information is delivered to the central
nervous system through
various sensory pathways in the form of spike trains. The pathways' information 
throughput has been studyed for a long time in theoretical \cite{MacKay0} and 
experimental \cite{Passaglia} research. The paradigm of those 
research is consistent with the assumption that the best situation is when the brain
receives maximum information from sensory systems. But 
compare this with \cite{Feldman}. Another paradigm, which is
as well mature, \cite{MacKay}, concentrates on self-organization of spike
trains when primary sensory activity spreads to higher
brain areas. Self-organization is accompanied with information loss, \cite{self},
Indeed, a kind of standartization of activity evoked by various primary sensory
inputs is obsereved experimentally in higher brain areas during 
olfactory \cite{Viret,Laurent} 
and auditory \cite{Cariani} perception. In visual system, simple examples are the
transformation of scene representation to viewpoint-invariant, \cite{Durbin}
or retinotopic-invariant \cite{Tang} coordinates. Here, spike trains in the optic
nerve must depend on certain information (retinal position, viewpoint), which is 
removed in higher areas of conceptual representation.
In the context of cognitive physiology, the process
of reduction of information aimed at conceptual representation/recognition of
external objects is known as information condensation, \cite{KonigKruger}.
Usually pattern recognition phenomenon, which is closely related to information 
condensation, is considered in parallel with 
training, see \cite{Adeli,Johnston0,Nichols0,Schliebs0,Soltic0,Strain0,Widrow},
learning, \cite{Acharya0,Adeli0,Adeli2,KonigKruger}, or other plasticity, \cite{Villa},
in the corresponding network.
In a biological network, the learning mechanism involves biosynthesis \cite{Scharf} 
and is therefore
very slow process, which requires seconds or minutes, \cite{Merzenich}.
At the same time,
recognition of objects in visual scene can be accomplished within 150 ms, or 
faster, cite{Thorpe1,Thorpe}. During this short period of time, the network has constant
structure, but it is the spiking activity which evolves within it from information-rich at the sensory periphery into 
information-poor, representing concrete 
entities/concepts at higher brain areas. 

\begin{figure*}
\begin{center}
\includegraphics[width=0.65\textwidth]{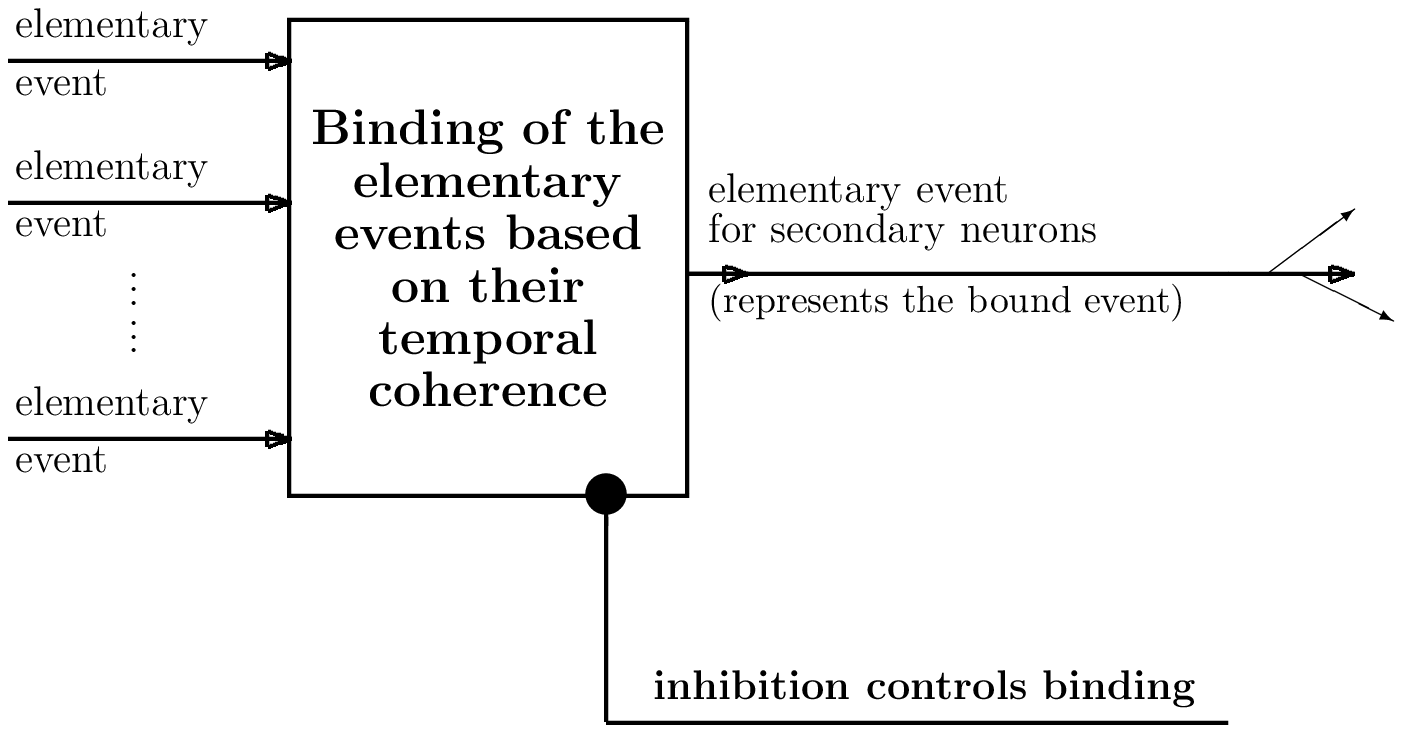}
\end{center}
\caption{Signal processing in the binding neuron model.\protect \cite{Vid96a,Vid98}}
\label{fig:paradigm}
\end{figure*}

We now put a question: What could be the physical mechanism of information
condensation in a spiking neural system? One possible mechanism, \cite{BNF}  
which operates at the level of single neuron, was discussed,
see Fig. \ref{fig:paradigm}, \ref{BNexample} 
and n. 2.1.1\ref{BNcond}, below.
In a reverberating spiking network, due to this mechanism 
the spiking dynamics should settle down to the same definite
periodic activity in response to any stimulus from a definite set of various inputs, 
and to another periodic activity in response to members of another set of inputs.
If so, then the definite periodic dynamical state can be considered as an abstract
representation of a feature, which all stimuli from the definite set have in common. 
And this is just what is expected from the condensation of information.

Our purpose in this work is to study how the ability of a simple spiking neural net 
to condense information in the above described sense depends on its physical
parameter --- the net's geometric size. For this purpose we simulate dynamics of
a net composed of five binding neurons placed equidistantly on a circle. The
circle's radius, $R$, characterizes the net's geometric size. The net is fully connected,
and propagation velocity is taken the same for all connections and all values of $R$.
Thus, the variations in $R$ are expressed exclusively  in the variations in the
interneuronal propagation times.
Initially, the net is stimulated by a spike train of input impulses triggering each of five
neurons at times $\{t_0,t_1,\dots,t_4\}$. Afterwards, the spiking  dynamics is
allowed to go freely until it 
settles down to a periodic one. This allows to figure out sets of input stimuli bringing
about the same periodic dynamics. The number of different periodic dynamics and
the number of input stimuli in a set, which corresponds to a definite periodic
dynamics, characterize the net's ability to condense information. We found that this
ability depends strongly on the  net's size.

\section{Methods}

\subsection{The Binding Neuron Model}
\noindent
The understanding of mechanisms of higher brain functions expects a continuous reduction from higher activities to lower ones, eventually, to activities in individual neurons, 
expressed in terms of membrane potentials and ionic currents. While this approach is correct scientifically and desirable for applications, the complete range of the reduction is unavailable to a single researcher/engineer due to human brain limited capacity. In this connection, it would be helpful to abstract from the rules by which a neuron changes its membrane potentials to rules by which the input impulse signals are processed into its output impulses. The “coincidence detector”, and “temporal integrator” are the examples of such an abstraction, 
see discussion by K\"onig et al., \cite{Konig}.
One more abstraction, the binding neuron (BN) model, is proposed 
as signal processing unit, \cite{Vid96} which can operate as either coincidence 
detector, or temporal integrator, depending on quantitative characteristics of 
stimulation applied. This conforms with behavior of real neurons, see, e.g. work by 
Rudolph \& Destexhe, \cite{Rudolph}.
The BN model describes functioning of a neuron in terms of discret events, 
which are input and output impulses, and degree of temporal coherence between 
the input events, see Fig. \ref{fig:paradigm}. Mathematically, this can be realized as
follows. Each input impulse is stored in the BN for a fixed time, $\tau$.
The $\tau$ is similar to the “tolerance interval” discussed by MacKay, \cite{MacKay}.
All input lines are excitatory and deliver identical impulses. 
The neuron fires an output impulse if the number of stored impulses, $\Sigma$, is equal or higher than the threshold value, $N_{0}$. In this model, 
inhibition is expressed in decreased $\tau$ value.
It is clear, that BN is triggered when a bunch of input impulses is received in a narrow temporal interval. In this case,
the bunch could be considered as compound event, and the output impulse --- as an abstract representation of this compound event. One could treat this mechanism as binding of individual input events into a single output event, provided the input events are coherent in time. Such interpretation is suggested by binding of features/events in largescale neuronal circuits, \cite{Damasio89,Eckhorn,Engel91a,Vid98}.

It would be interesting to characterize the BN input-output relations in the form of transfer function, which allows exact calculation of output in terms of input. 
In our case, input is the sequence of discrete arriving moments of standard
impulses: $T_{in}=\{l_1,l_2,l_3,l_4,\dots\}.$	
The output   is the sequence of discrete firing moments of BN: $T_{out}=\{f_1,f_2,\dots\}.$	
It is clear that $T_{out}\subset T_{in}.$ The transfer function in our case could be the function $\sigma(l)$, $l\in T_{in}$, which equals 1 if $l$  is the firing moment, $l\in T_{out}$, and 0 otherwise. 
For BN with threshold $N_0$ the required function can be constructed as follows.
It is clear that the first $N_0-1$ input impulses are unable to trigger neuron, therefore $\sigma(l_1)=0,\dots,\sigma(l_{N_0-1})=0.$	
The next input is able to trigger if and only if all $N_0$  inputs are coherent in time: 
$$\sigma(l_{N_0}) =1 \textrm{ if and only if } l_{N_0}-l_1\le\tau.$$
In order to determine $\sigma(l_{N_0+k})$, $k\ge1$, one must take into acount all previous input moments, therefore we use notation $\sigma_{T_{in}}$ instead of $\sigma$. The values of $\sigma_{T_{in}}(l_{N_0+k})$ can be determined recursively:
\begin{gather}
	\sigma_{T_{in}}(l_{N_0+k})=1\textrm{ if and only if }\nonumber
	l_{N_0+k}-l_{k+1}\le\tau\textrm{ and }\nonumber
\\\nonumber
	\sigma_{T_{in}}(l_i)=0\textrm{ for all } i\in\{k+1,\dots,N_0+k-1\}.
\end{gather}
The function $\sigma_{T_{in}}$  describes completely the BN model for arbitrary threshold value $N_0\ge2$ .

\subsubsection{Information Condensation in a Single Neuron}\label{BNcond}

 \noindent
It is worth noticing, that any firing (triggering)  moment of a spiking neuron 
is determined by the moment of last input impulse, which just
ensures that the triggering condition is satisfied. In a neuron, which needs more
than one input impulse to fire, variations of temporal position of impulses,
received just before the triggering one, do not influence the moment of emitting
the output impulse, provided those variations are in resonable limits and 
arrival moment of the triggering impulse remains the same.\smallskip

\begin{figurehere}
\begin{center}
\includegraphics[width=0.4\textwidth]{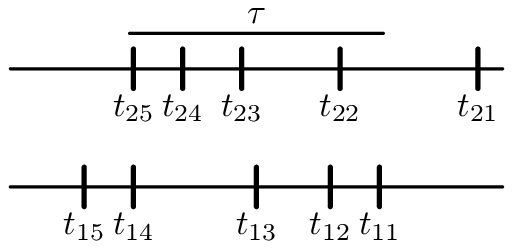}
\end{center}
\caption{\label{BNexample}Example of two different inputs into a single neuron, which produce identical outputs.}
\end{figurehere}\smallskip

\noindent
Thus, different input
spike trains can produce exactly the same output. This looks like if some
detailes of the input stimulus, which is composed of several impulses,
were reduced/condensed in the output.
In the BN, the triggering condition is
that the number of impulses in the BN's internal memory equals to $N_0$.  
Consider BN with $N_0=4$, which is stimulated with two different spike trains
with input impulses arrival times $S_1=\{t_{11},t_{12}, t_{13}, t_{14}, t_{15}\}$,
$t_{11}<t_{12}< t_{13}< t_{14}< t_{15}$, and 
$S_2=\{t_{21},t_{22}, t_{23}, t_{24}, t_{25}\}$, 
$t_{21}<t_{22}< t_{23}< t_{24}< t_{25}$. Let the arrival moments satisfy the
following conditions:
\begin{gather}\label{exampleBN}\nonumber
t_{14}-t_{11}<\tau, \quad t_{25}-t_{22}<\tau,
\\\nonumber
t_{24}-t_{21}>\tau,\quad t_{14}=t_{25}=t_o\,.
\end{gather}
In this case, both $S_1$ and $S_2$, if fed to the BN, will produce exactly the same output,
namely, the single impulse at moment $t_o$. This is illustrated in 
Fig. \ref{BNexample}.

\subsection{The Network Model}\label{Network}

\begin{figure*}
\begin{center}
\includegraphics[width=0.55\textwidth]{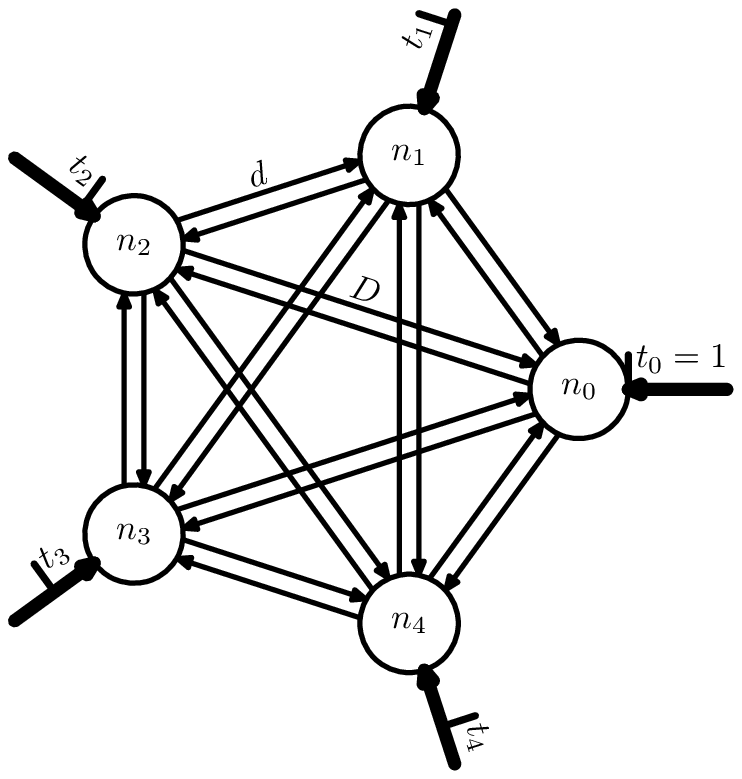}
\end{center}
\caption{The network, used for simulations. Here $\{t_0,t_1,t_2,t_3,t_4\}$ --- is the input spike train, $d$, $D$ --- are the propagation delays in the connection lines. Any line can be either empty, or propagating one impulse.}
\label{net}
\end{figure*}

\noindent
As a reverberating spiking neural net we take the net of five 
neurons placed equidistantly at a circle of radius $R$, see Fig, \ref{net}. 
Each neuron has threshold $N_0=4$, and internal memory, $\tau=10$ ms.
The net is fully connected. The connection lines are characterized with
length and propagation velocity, $v$, which is the same in all lines. 
For $R$ fixed, there are two types of connection line, the short one,
with propagation delay $d$, and the long one with propagation delay $D$. 
Each neuron has additionally the external stimulus input line, which is used to 
start the net dynamics. Single impulse in the stimulus input line delivers to its
target neuron just threshold excitation. This causes firing at the moment
of the stimulus impulse arrival.
For numerical simulations we use 20 networks of different sizes, see Table \ref{T1}. 
The propagation
velocity in any interconnection line is taken $v=0.1$ m/s.

\subsubsection{Numerical Simulation}
\noindent
As programming language we use Python under Linux operating system.
The dynamics was modelled by advancing time with step $dt=200$ $\mu$s.
The delay values $d$ and $D$, when measured in the $dt$ units,
were rounded to the nearest from below
integers, see Table \ref{T1}. As a result, the simulating program operates in whole numbers 
with no rounding errors involved. 
\begin{table*}
\begin{center}
\begin{tabular}{c|cccccccccccccccccccc}
net \# & 1&2&3&4&5&6&7&8&9&10\\
\hline
$R$, mm& 0.029 & 0.057 & 0.086 & 0.114 & 0.143 & 0.171 & 0.200 & 0.229 & 0.257 & 0.286 \\
\hline
$d$, $dt$ & 1 & 3 & 5 & 6 & 8 & 10 & 11 & 13 & 15 & 16\\
\hline
$D$, $dt$ &2 & 5 & 8 & 10 & 13 & 16 & 19 & 21 & 24 & 27\\
\hline
$M$, $dt$ &5 & 10 & 15 & 20 & 25 & 30 & 35 & 40 & 45 & 50\\
\hline
\end{tabular}

\begin{tabular}{c|cccccccccccccccccccc}
\hline
net \# &11&12&13&14&15&16&17&18&19&20 \\
\hline
$R$, mm&  0.314 & 0.343 & 0.371 & 0.400 & 0.429 & 0.457 & 0.486 & 0.514 & 0.543 & 0.571\\
\hline
$d$, $dt$ & 18 & 20 & 21 & 23 & 25 & 26 & 28 & 30 & 31 & 33\\
\hline
$D$, $dt$ & 29 & 32 & 35 & 38 & 40 & 43 & 46 & 48 & 51 & 54\\
\hline
$M$, $dt$ &55 & 60 & 65 & 70 & 75 & 80 & 85 & 90 & 95 & 100
\end{tabular}
\end{center}
\caption{\label{T1}Dimensions of networks used for simulations; $dt=200$ $\mu$s.}
\end{table*}

\begin{table*}
\begin{center}
\begin{tabular}{ccccccccccccccccccccc}
net \# & 1&2&3&4&5&6&7&8&9&10\\
\hline
 & 3$^1$ &6$^1$ &9$^1$ &11$^1$&14$^1$&17$^1$&20$^1$&22$^1$&25$^1$&28$^1$\\
 &  &10$^4$&15$^4$&18$^4$&23$^4$&28$^4$&32$^4$&36$^4$&41$^4$&45$^4$\\
&  &12$^3$ &18$^7$&22$^7$&28$^7$&34$^7$&40$^7$&44$^7$&50$^7$&56$^7$\\
&  &     &24$^6$&29$^6$&37$^6$&45$^6$&52$^6$&    &  &\\
\hline
\end{tabular}

\begin{tabular}{ccccccccccccccccccccc}
\hline
net \# &11&12&13&14&15&16&17&18&19&20 \\
 \hline
 & 30$^1$ &33$^1$ &36$^1$ &39$^1$&41$^1$&44$^1$&47$^1$&49$^1$&52$^1$&55$^1$\\
 & 49$^4$ &54$^4$ &58$^4$ &63$^4$&    &    &    &    &    &    \\
 \end{tabular}
\end{center}
\caption{\label{T2}Distinct periods in $dt$ units of found periodic states in the short stimuli paradigm. Superscript denotes the number of different periodic states with this period.}
\end{table*}

Each single tick of the program, the network's tick,  advances time by $dt$, and consists of three partial ticks, which are performed in the given order. Namely, (i) input tick, which advances time in the input lines, (ii) axonal tick, which advances time in the internal connection lines, (iii) neuronal tick, which advances time in the neurons. 

This manner of updating 
states can be treated as synchronous in a sense that each component of the
network has the same physical time when the network's tick is complete. On the other hand, 
viewed as interneuronal communication process, the dynamics should be treated as 
asynchronous due to nonzero propagation delays.

During the step (ii), a neuron can get impulse into its internal memory. 
If a neuron appears in the state 
``Fire'' as a result of the network tick, then the output impulse it produces can
appear in the connection lines only during the next network tick. This introduces
effective delay of one $dt$ between delivering the triggering impulse to a
neuron and emitting output impulse by that neuron.

\subsection{Data Acquisition Algorithm}
\subsubsection{Set of Stimuli}
\noindent
The net was entrained to reverberating dynamics by applying initial input
spike train of five impulses, one triggering impulse per neuron, at times
(in $dt$ units) 
$\{t_0=1,t_1,t_2,t_3,t_4\}$. The triggering moment of neuron \# 0 is taken
1 for all stimuli in order to exclude rotational symmetry between the stimuli applied. 
Other four triggering moments run independently through the set
$\{1,2,\ldots,t_{max}\}$, where $t_{max}$ is choosen proportional to $R$
for each net size. In choosing $t_{max}$, we follow two different paradigms.
In the first paradigm of short stimuli we restrict the overall duration of the stimulus train
with the value $t_{max}=d$. Thus, the set of stimuli has $d^4$ different stimuli.
If $t_i\le d$, $i=0,\dots,4$, then any neuron in the net never obtains
impulse from other neurons before it obtains its external input stimulation.
In the second paradigm of extended stimuli, we restrict the overall duration of the stimulus train
with the value $t_{max}=M$, which is about three times longer than $d$ for each 
network (see Table \ref{T1}).
Here, all stimuli $\{t_0=1,t_1,t_2,t_3,t_4\}$, which were presented to a network, 
cover the set of $M^4$ different trains, which equals 
from 625 different stimuli for net \#1 to 100\,000\,000 different
stimuli for net \#20 (see Table \ref{T1}). The stimuli were sampled in accordance
with standard algorithm of 4-digit counter. Namely, we started from stimulus
$\{1,1,1,1,1\}$, the next stimulus is obtained by advancing $t_1$ by 1, and so on.
The stimulus next to $\{1,M,1,1,1\}$ is $\{1,1,2,1,1\}$, the one next to $\{1,M,M,1,1\}$ is $\{1,1,1,2,1\}$ and so on, until stimulus $\{1,M,M,M,M\}$ is presented.

In the extended paradigm, the late external input impulse can enter corresponding neuron
after it received impulses from neurons already triggered by early external input impulses.

The second paradigme is in concordance with visual information processing, \cite{Bullier}
where activity from higher brain areas, which was invoked due to visual stimulation
at earlier time, 
is retroinjected to areas V1 and V2 in the primary visual cortex, where it interacts with activity invoked by visual
input at later time during perception.

\subsubsection{Figuring out Periodic States}
\noindent
After the last input impulse from the train $\{t_0=1,t_1,t_2,t_3,t_4\}$ reachs its target neuron,
the program begins appending at each time step the instantaneous state of the net to a Python list.
The instantaneous state consists of states of all 20 connection lines and all 5 neurons
 (see Fig. \ref{STATES}).
Before appending, the program checks if the current instantaneous state was already 
included in the list. If it was, then the periodic dynamical state is found with its complete cyclic trajectory covered by instantaneous states between 
the inclusion and the last record in the list, inclusively.
Measures are taken in order not to count the same cyclic trajectory, which was
entrained at its different points, as different periodic states.

\begin{figure*}
\begin{verbatim}
mysql>  select * from STATES_5_9 where num=269;
+------+------------------------------------------------------------------------------
----------------------------------------------+--------+
| num  | state                                                                        
                                              | period |
+------+------------------------------------------------------------------------------
----------------------------------------------+--------+
|  269 | 0 1 1 0 8 8 17 17 24 15 15 24 8 8 0 0 8 17 17 8 
0 False False 
0 False False 
0 False False 
0 False False 
0 False True 49 1 
 |     82 | 
+------+------------------------------------------------------------------------------
----------------------------------------------+--------+
1 row in set (0.02 sec)

mysql> 
\end{verbatim}
\caption{\label{STATES}Example of single record in the MySQL table STATES. 
The first field (num) is numerical, and gives the serial number of periodic state found.
The second field (state) is a string, which describes instantaneous state from the 
periodic state found (a point from the cyclic trajectory, which represents the whole trajectory, or periodic state). The first 20 numbers in the string describe states of all 
connection lines: '0' means that the line is empty, positive number specifies after 
how many ticks the propagating impulse will rich the targeted neuron. The next five 
chunks confined between the "next line" symbols describe states of neurons. 
The first number in each chunk is the "kick" --- the total number of impulses obtained
by neuron after the axonal tick was complete. During the neuronal tick, corresponding 
to that axonal tick, the "kick" is utilized and set to zero. The next boolean in the 
chunk indicates if the neuron is in the "Fire" state. The next boolean indicates if the 
neuron has any impulses in its internal memory. If it has, then next couples
of numbers (up to three couples) describe those impulses. In this example,
neuron \#4 has 1 impulse with time to leave 49$\cdot dt$. The third field (period)
specifies period (in $dt$ units) of this periodic state.}

\end{figure*}

\medskip 
\begin{figure*}
\begin{center}
\includegraphics[angle=-90,width=0.45\textwidth]{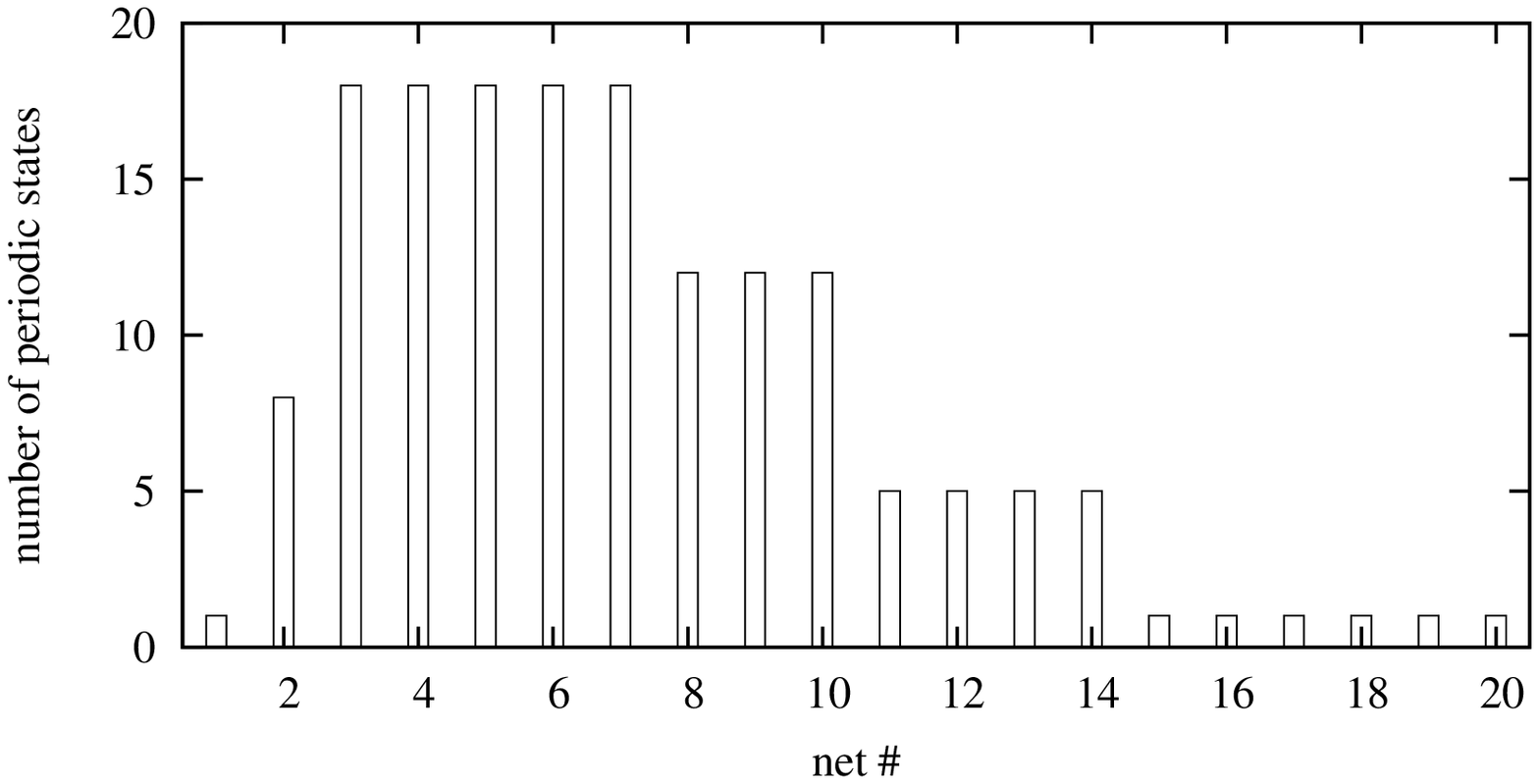}
\hfill
\includegraphics[angle=-90,width=0.45\textwidth]{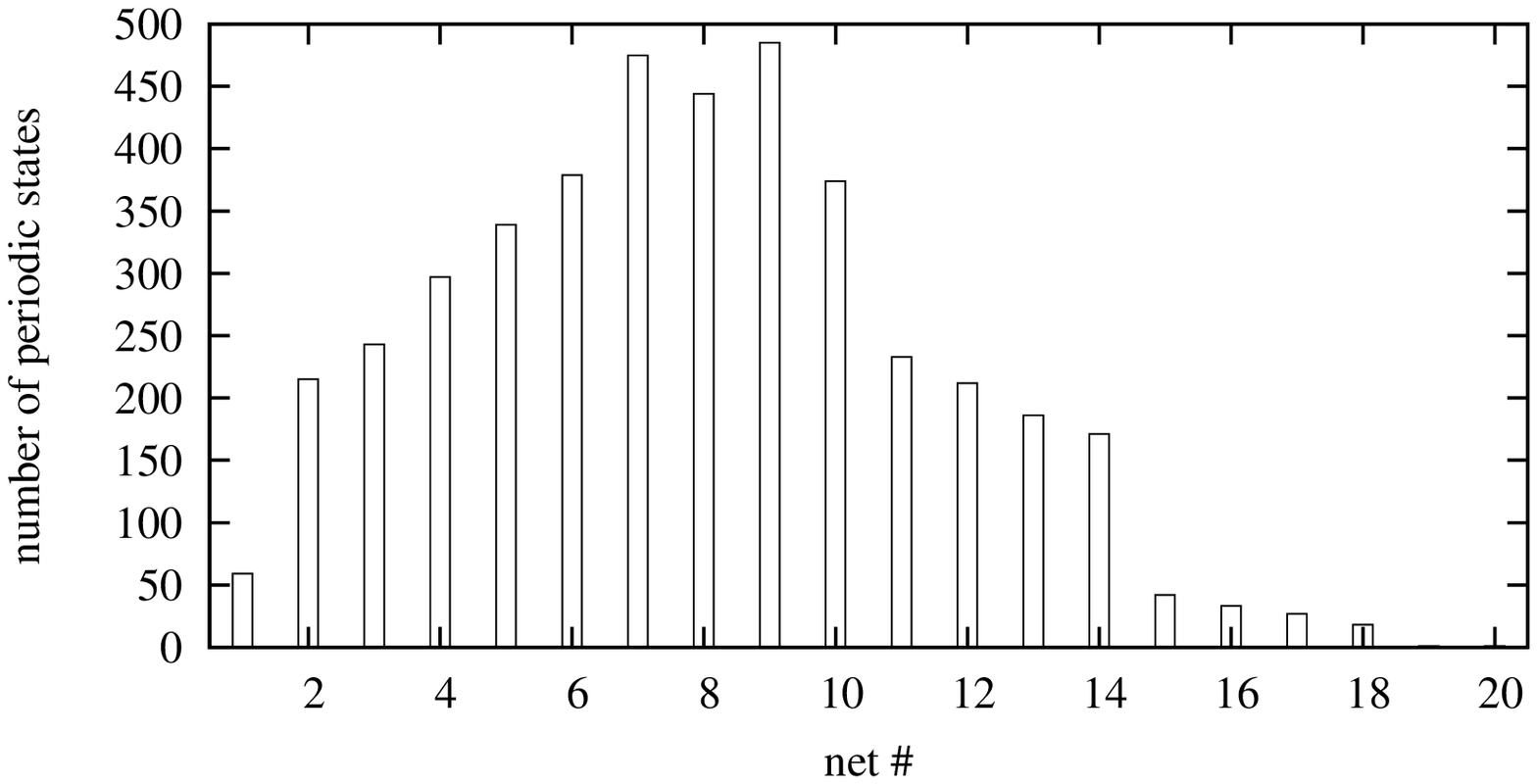}
\end{center}
\caption{\label{NS}Number of different periodic states found for each 
net with short (left panel), and extended (right panel) stimuli.}
\end{figure*}\medskip

The data for each net were stored in two MySQL tables. Table STATES included 
the serial number of 
any periodic state found, one element from the corresponding cyclic trajectory, and
period of the state (see Fig. \ref{STATES}). Single record in the INPUTS table included the input stimulus $\{t_0=1,t_1,t_2,t_3,t_4\}$,
the serial number of the periodic state it leads to (this number is 0 for fading dynamics), and relaxation time, namely, the time,
which is spent between the last external input impulse is delivered and the net's
entrance moment into the periodic regime.

\section{Results}
\subsection{Characterization of Periodic States Found}
\noindent
After initial stimulation, networks from \#1 to \#7 entrain to periodic activity
after any stimulation, and networks \#8 to \#20 either entrain to periodic dynamics,
or stops from any activity after some time. This takes place for both short and extended stimuli.The number of different periodic states found for each network is shown in
Fig. \ref{NS}. Here, the maximal number of periodic states obtained with short stimuli
is 18, and this number is achieved for net numbers from 3 to 7. Exact values of periods, and number
of different periodic states with this period is shown in Table \ref{T2} for short stimuli. We omit similar table for extended stimuli.
The maximal number of different
periodic states obtained with extended stimuli is 485, which is achieved in the net number 9.
The maximal number of different periodic states with the same period is 294 for period 50$\cdot dt$ in net \#9. It should be mentioned that two periodic states,
which can be turned into eachother by suitable renumeration of neurons, were
considered as different.

It is evident that in the network of five neurons with threshold 4, each neuron is triggered the same number of times during period. Indeed, expect that neuron $n_4$
fires $k_4$ times, and any of the other four fires less during period: 
$k_4>k_i,~i=0,1,2,3.$ In order to be triggered $k_4$ times, $n_4$ must obtain not less than $4k_4$ input
impulses during period. But it can obtain only $k_0+k_1+k_2+k_3$, which is less than
required. Similarly, situation when $k_4=k_3$ and $k_3>k_i,~i=0,1,2$ leads to contradiction, and so on.

This number of triggering is
either 1 or 2 for trajectories found, see examples in Fig. \ref{NETexamples}. Some nets have only one periodic state, which coresponds to synchronous firing of all 5 neurons
and symmetrical states of connection lines at any moment of time. 
This is the case for nets number 1 and from 15 to 20
for short stimuli, and for nets number 19 and 20 for extended stimuli.

\subsection{Condensation of Information}\label{Coi}
\noindent
In order to estimate the degree of information condensation in the course of 
transformation of an external spike train into a certain periodic state of the net, one needs to calculate
information amount, which is delivered by specifying a spike train, and which is delivered by specifying the state, it leads to.\smallskip

\begin{figurehere}
\begin{center}
\includegraphics[width=0.4\textwidth]{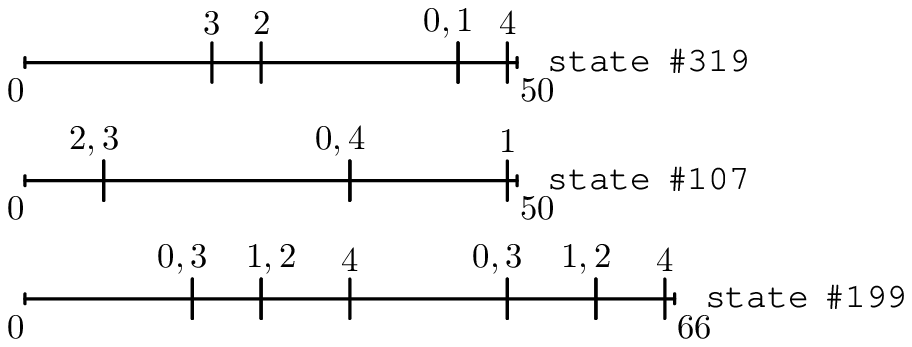}
\end{center}
\caption{\label{NETexamples}Examples of periodic states found for net number 9 
in the extended stimuli paradigm. Spikes
indicate the firing moments, labels near each spike give numbers of neurons, 
firing at this moment. The two upper trains show states with period 10 ms, the lower
one - with period 13.2 ms.}
\end{figurehere}\smallskip

\noindent
 This can be done by wellknown Shannon's
formula \cite{Shannon}
\begin{equation}\label{Sh}
H=-\sum_i p_i \log_2 p_i,
\end{equation}
where $p_i$ is the probability to obtain case number $i$ from a set of cases. At the 
input end we have the set of $d^4$, or $M^4$ different external input spike trains.
In our statement of the problem, it is natural to consider all external input trains as
equally probable. If so, then information delivered by specifying certain train is
\begin{equation}\label{Hs}
H_s=4\log_2 d,
\end{equation}
for the short stimuli paradigm, and
\begin{equation}\label{He}
H_e=4\log_2 M,
\end{equation}
for the extended stimuli paradigm.

While estimating information, delivered by specifying certain periodic state, one 
should take into account that probabilities of different periodic states are not the same.
In order to calculate probability $p_n$ of a periodic state $C_n$,
we calculate the number of
input spike trains leading to the $C_n$, namely, $T_n$, and divide this number 
by the total amount of different input stimuli (see histogram for $T_n$ in 
Fig. \ref{Histo})
\begin{equation}\label{Ps}
p_n=\frac{T_n}{d^4},
\end{equation}
for the short stimuli paradigm, and
\begin{equation}\label{Pe}
p_n=\frac{T_n}{M^4},
\end{equation}
for the extended stimuli paradigm. Then we use Eq. (\ref{Sh}) with probabilities
of individual periodic states found in accordance with (\ref{Ps}), (\ref{Pe})
to calculate information
which should be ascribed to any periodic state. 
In this calculations, we treat
uniformly with others the external input stimuli, which lead to fading dynamics.
Correspondingly, the state with no activity is treated uniformly with periodic states.
This is in the contrast with data presented in Fig. \ref{NS},
and Table \ref{T2}, where the state with no activity is excluded.\smallskip

\begin{figurehere}
\begin{center}
\includegraphics[angle=-90,width=0.45\textwidth]{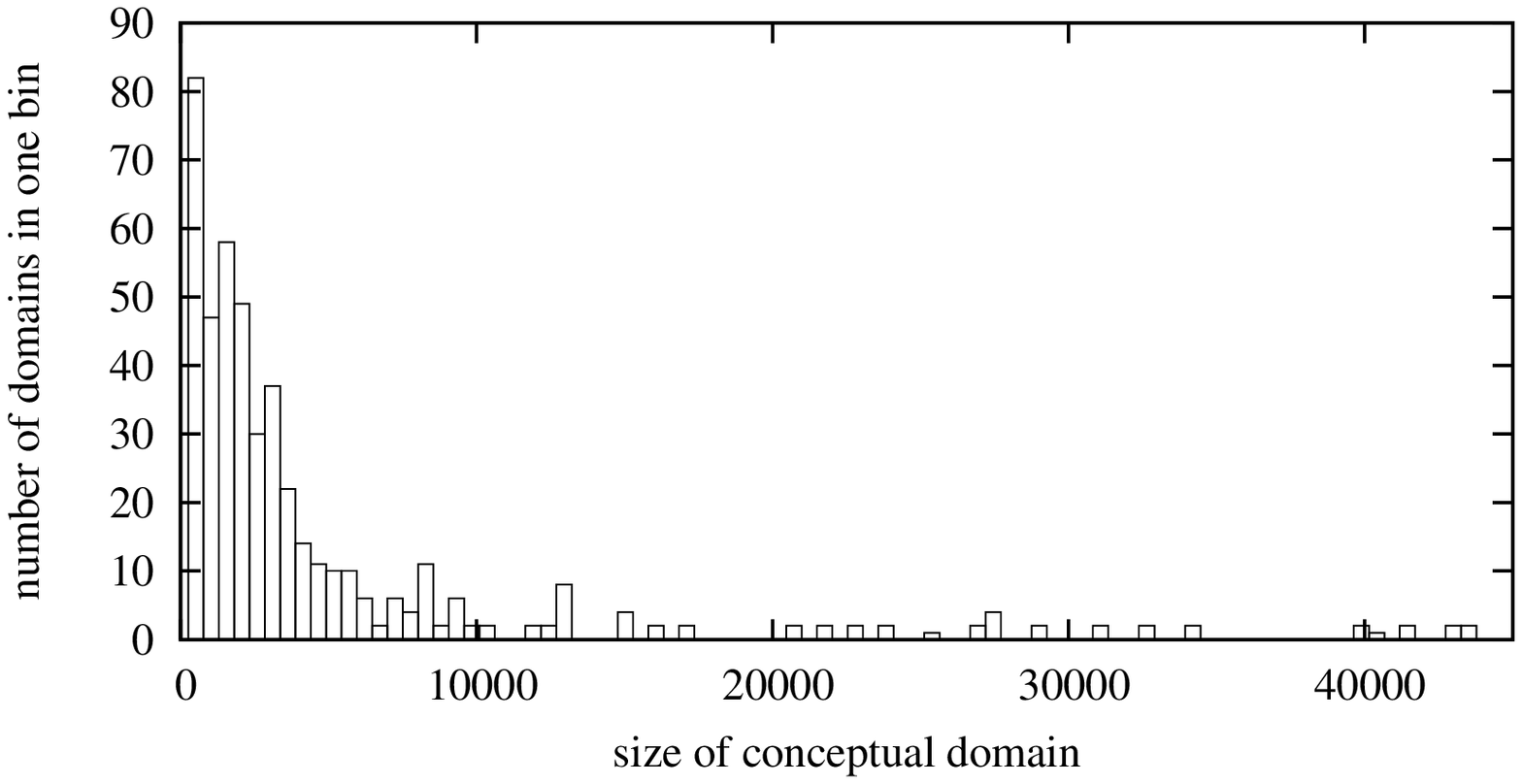}
\end{center}
\caption{\label{Histo}Histogram of conceptual domain sizes for net \#9 in extended stimuli paradigm. The bin size is 518. 23 domains with sizes from 50764 to 193732 are not presented.}
\end{figurehere}\smallskip

As it could be expected, the information amount in an external input spike train
increases as logarithm of the net size, in accordance with Eqs. (\ref{Hs}), 
(\ref{He}), varying from 25 to 81 bits for short paradigm, and from 37 to 106 bits for
extended paradigm.
Information, which could be ascribed to a periodic state, depends on the net size
in a more complicated manner, see Fig. \ref{info}. A remarkable feature is a kind 
of plateau between net \#3 and \#9 for both short, and extended stimulation 
paradigme. In the plateau, the\smallskip

\begin{figurehere}
\begin{center}
\includegraphics[angle=-90,width=0.45\textwidth]{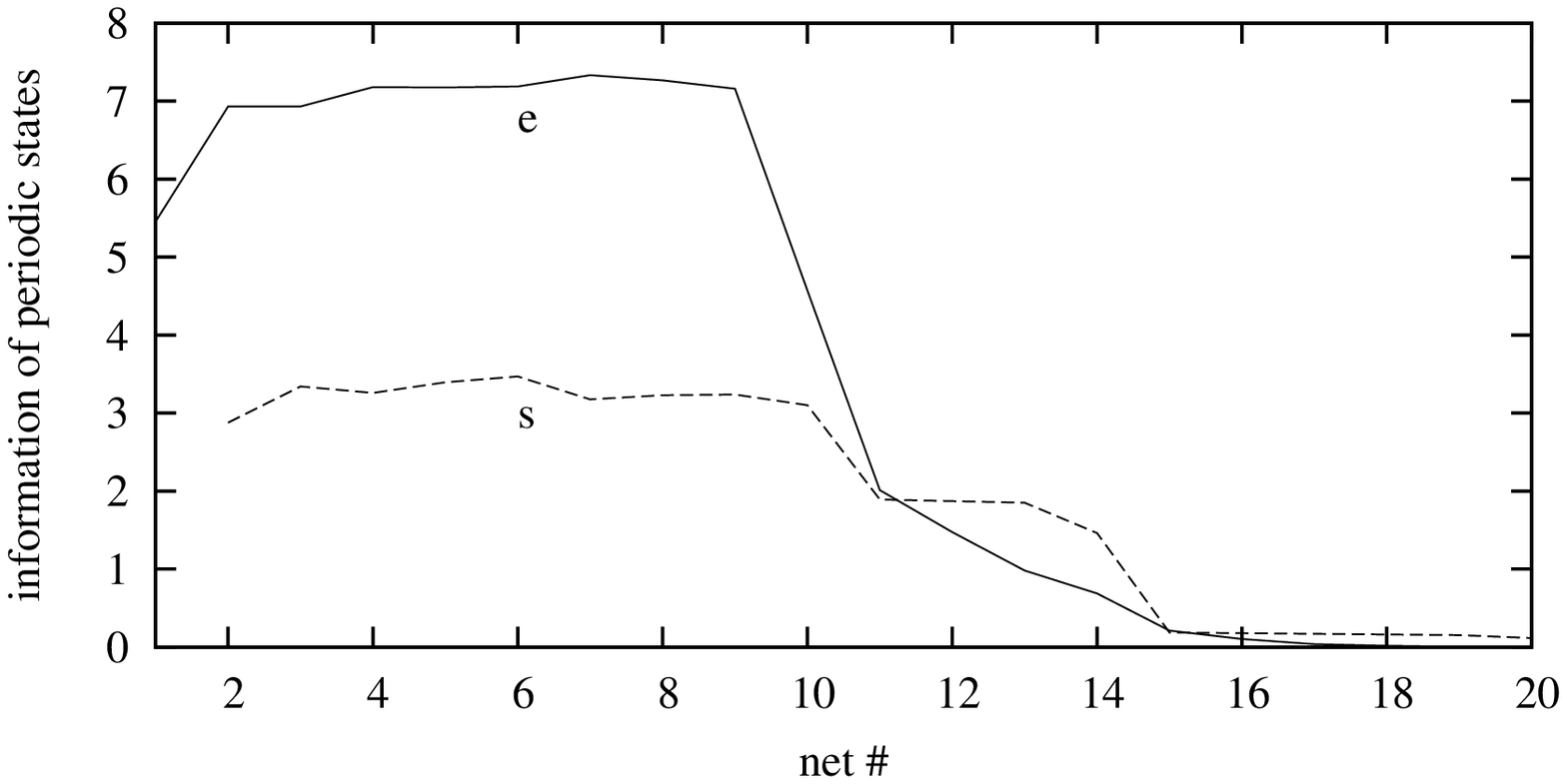}
\end{center}
\caption{\label{info}Dependence of information amount, which is ascribed to periodic state, on the net size. Curve 'e' corresponds to extended stimuli paradigm, 's' --- to the short stimuli one.}
\end{figurehere}\smallskip

\noindent
information of a periodic state, calculated in
accordance with Eq. (\ref{Sh}) with probabilities found due to Eqs. 
(\ref{Ps}), (\ref{Pe}), varies  between 6.93 and 7.33 bits for extended stimuli,
and between 3.17 
and 3.46 for short stimuli. In the plateau, the degree of information condensation
calculated as input information divided by the periodic state information, varies
between 9.02 and 12.27 for extended stimuli, and between 11.2 and 19.31 for
short stimuli. 
Out of a plateau range, for greater net sizes the amount of information in a periodic
state drops sharply due to simplification of the set of periodic states. Namely,
the total number of periodic states decreases to one, with the second one with no 
activity, and the probability is distributed very
unevenly between this two states. This leads to the degree of information 
condensation as high as 41000 for extended stimuli, and 690 for short ones.

\section{Discussion}

\noindent
Any reverberating spiking neural net can represent complicated dynamical behavior. 
If the net's instantaneous states can be described with whole numbers, then the net 
will inevitably either entrain to periodic dynamics, or stop its activity at all. 
In this study, it is appeared
that a very simple net of Fig. \ref{net} can be engaged into a considerably large set 
of different periodic activities. It is not clear which part of all possible
in this net periodic states was discovered in our simulation. As it follows
from comparison between short and extended stimuli paradigm,
the number of different periodic states found increases with 
increasing range of input stimuli. Certainly, this increase must saturate
somewhere. This is because any two different periodic regimes are represented
by their cyclic trajectories, which has no common points (instantaneous states).
On the other hand, the total number of instantaneous states the network can have
is finite due to finiteness of the set of states of each element the network
is composed of.

The number of periodic states found in a net depends on the net's geometric size.
Variations in the net's 
size display themselves exclusively in variations of interneuronal propagation 
delays $d$ and $D$. 
On the other hand, the duration of neuronal internal memory, $\tau$,
is the same for net of any size. Thus, it is namely the relationships between 
the times an impulse spends for travelling between neurons, and time
it is allowed to spend in a neuron waiting for additional impulses, 
which controls possible number of periodic states.

It is worth noticing that each net has one completely synchronized  periodic state.
The completely synchronozed state is stable and achieved during finite time.
This is in the contrast to the case of pulse-coupled oscillators with delayed
excitatory coupling, (see, e.g. Ref.  \cite{Wu}).

In the four-dimensional set of stimuli we used, the neighbouring stimuli differ
from eachother by one $dt$ in one of four dimensions. This can be treated as
analogous representation of some reality. The set of periodic states
should be considered as a set of discrete entities due to
qualitative difference between any two states. This conforms
with a paradigm discussed in cognitive physiology, \cite{KonigKruger}.
The process
of transformation of initial analogous inputs into a discret set of periodic
states implies a loss of information and can be treated as 
condensation of information.

If we take a set of input stimuli, any of which leads to the same periodic state,
then that periodic state can be considered as an abstract/conceptual representation
of a feature, which all stimuli from the set have in common, 
and the corresponding set could be named as ``conceptual domain''.
What kind of feature or concept does the conceptual domain represent?
 If our net was trained to recognize a certain real feature, then it would be
that feature. In the context of this study, the common feature is that all stimuli
from the conceptual domain engage namely this net into  namely this
 periodic dynamics.  \smallskip

\begin{figurehere}
\begin{center}
\includegraphics[width=0.2\textwidth]{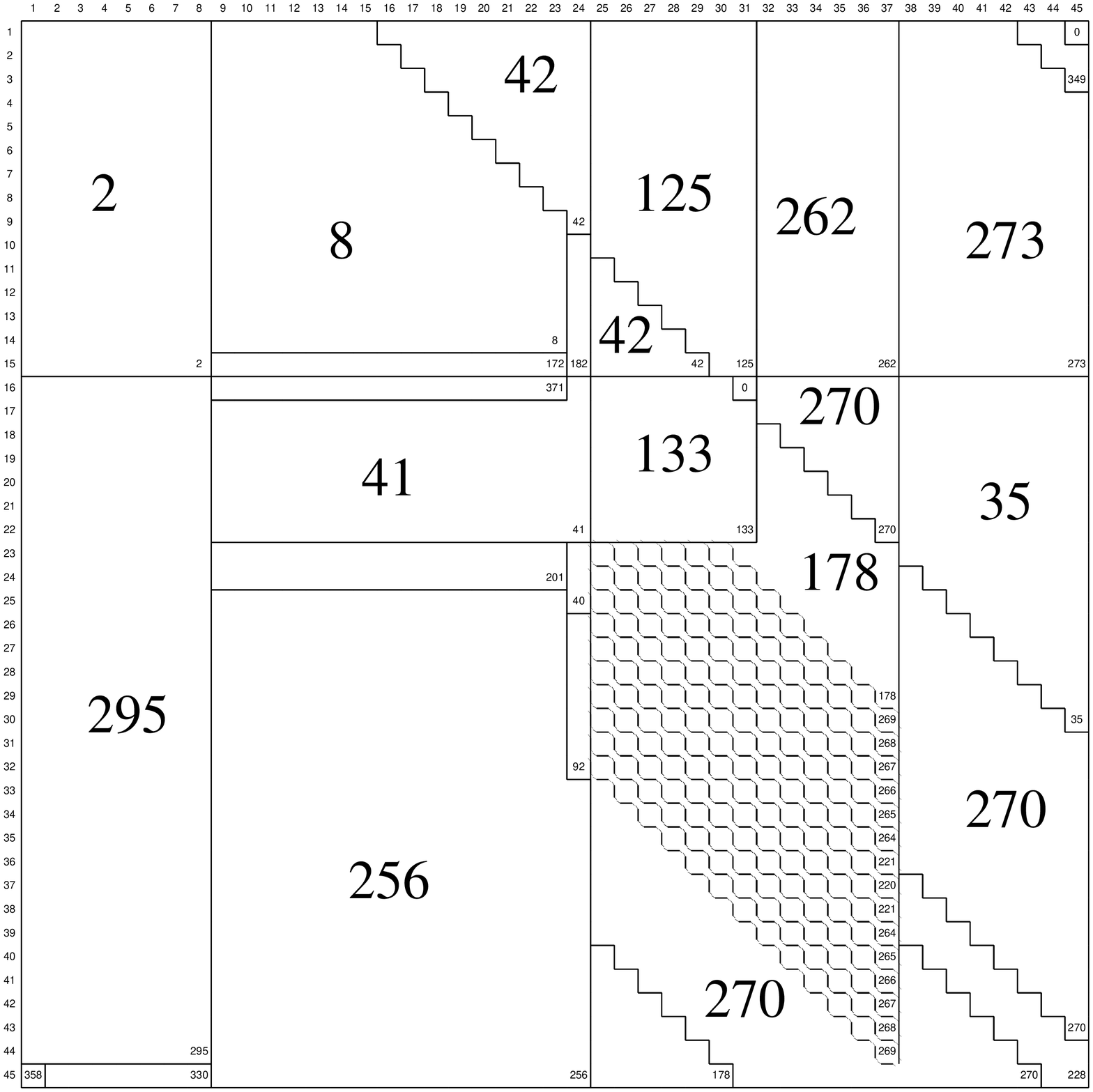}\hfill
\includegraphics[width=0.2\textwidth]{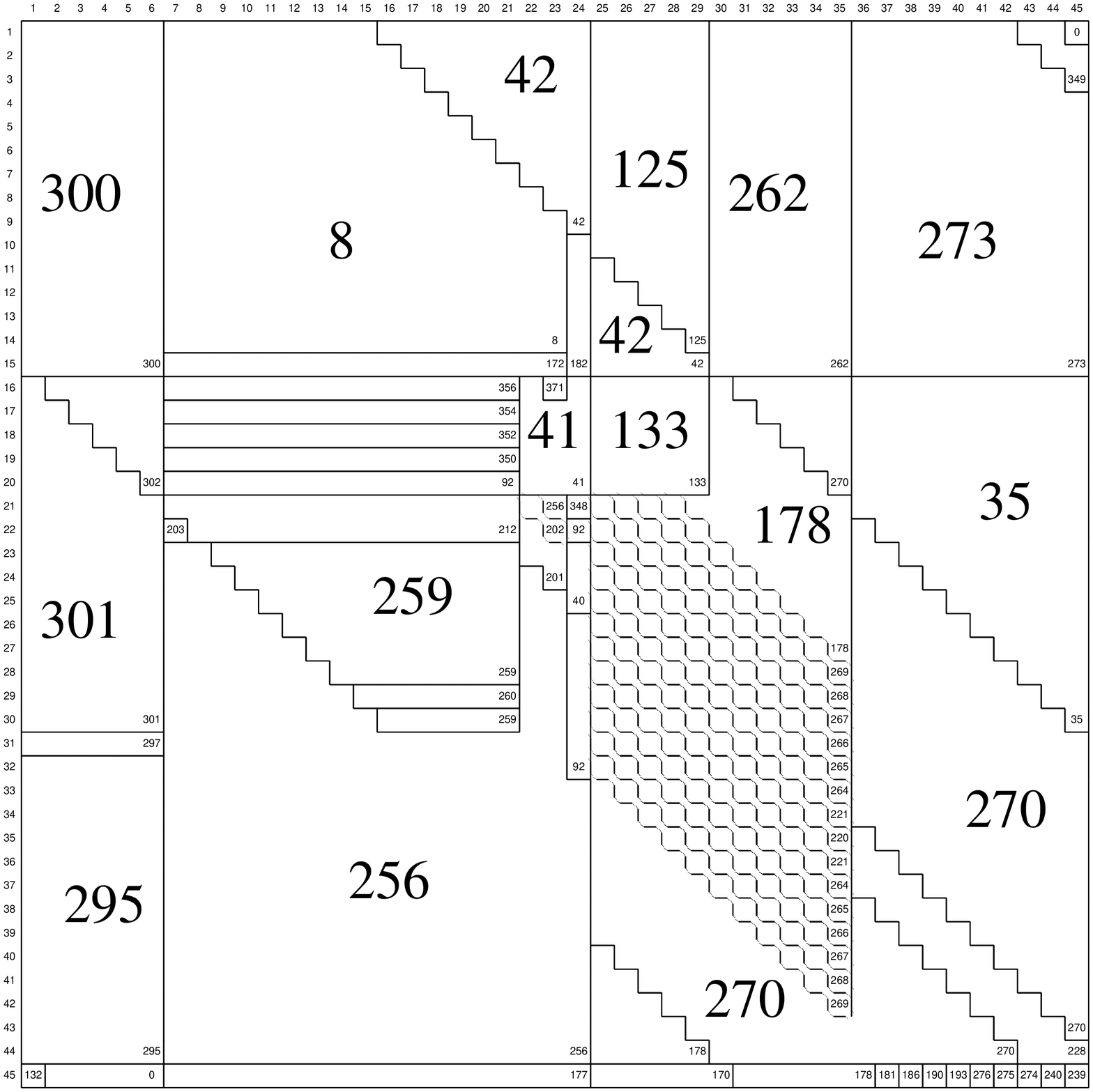}\hfill
\end{center}
\caption{\label{domains}2-dimensional cross-section of the 4-dimensional space of inputs for network \#9 in extended stimuli paradigm. {\it left,} cross-section is made
with plane $(t_3=23, t_4=23)$, {\it right,} $(t_3=21, t_4=23)$.  Origin for $(t_1,t_2)$
is in the upper left corner. Both $t_1$ and $t_2$ run through the set 
$\{1,2,\dots,45\}$ of values. Numbers in polygons indicate serial numbers of corresponding periodic states.}
\end{figurehere}\smallskip

It would 
be interesting to have a look on the topology of the conceptual domains in the
4-dimensional space of all stimuli. For this purpose we figured out 2-dimensional
cross-section of the input stimuli (see Fig. \ref{domains}). In the cross-section,  a
typical conceptual
domain is composed of several coherent clusters disconnected with eachother.
The histogram of sizes of conceptual domains found  is given 
in Fig. \ref{Histo}. 

Decomposition of the whole set of input stimuli into a number of conceptual domains
represented by corresponding periodic dynamics resembles approach in analyses
of multivariate datasets, see review in \cite{Gorban}. The difference is that here
the 4-dimensional dataset (a conceptual domain) is represented by unidimensional
cyclic trajectory, which corresponds to the domain, and the trajectory is composed 
of points/vectors, which have other dimension than the dataset vectors
(see Fig. \ref{STATES}). Nevertheless, having the
network, the whole cyclic trajectory can be reproduced starting from its 
any single point. Thus, here the datasets are reduced down to individual points.
This is in concordance with the information condensation idea.

Why do we stick ourselves with namely the periodic dynamics? The answer is
related to the memory/learning problem, even if we do not consider any plasticity 
in this study. It is known \cite{Merzenich} that modification of synaptic strength may happen
due to repetitive delivery of impulses to those synapses. Periodic dynamical states 
are just well suited for such repetitive delivery. All other dynamical behaviors are of
transient type, and have less chances to cause plastic changes in biological
network. On the other hand, successful perception expects ability to report 
about what was perceived, which is impossible without memory.

Unfortunately, we cannot draw this biological analogy too far. Real biological network
includes not only excitatory, but about 4\% of inhibitory neurons, \cite{Schwartz}.
Representing this characteristic in a model network requires to have at least 25
neurons in it. This dramatically increases the number of possible external
stimuli, which requires a qualitative change in our approach as regards the speed
of simulations and analysis of obtained data.

Finally, what happens if we use another neuronal model in the network? Our opinion
is that results will be qualitatively similar. Using the binding neuron here is natural,
since it represents in refined form what a spiking neuron does with signals
it receives. Additionally, the BN model easily allows to develop a program
operating in whole numbers. This excludes possible dynamical artefacts due to 
rounding errors.

In future work, it would be interesting to compare results, if another spiking neuron
model is used in the network, to study the topology of conceptual domains and
how the topology could change if a plasticity is introduced in the network model.

\section{Conclusions}

\noindent
A network composed of spiking neurons is able to condense information due to the 
fact that different initial stimuli could lead the network to the same periodic dynamics.
This happens by means of initial/basic condensation of information in spiking neurons, as it is described in n. 2.1.1\ref{BNcond}, above. The network's geometric size,
which determines the interneuronal transmission delays, has considerable influence
on the net's ability to condense information, mainly due to influence on the number 
of different periodic states the network can have, see Fig. \ref{NS}. The latter has 
influence on the amount of information, which should be ascribed to a single
periodic state, see Fig. \ref{info}. As a result, the degree of information condensation
varies between 9 and 41\,000, see n. 3.2\ref{Coi} for details.

The networks considered here are too primitive to have reliable biological implications.
At the same time, numerical parameters, see Table \ref{T1} and n. 2.2\ref{Network},
such as network sizes and spike propagation velocity, are taken corresponding to
biological data. The threshold value 4 does not contradict to biological reality, as
experimentally registered thresholds are between 1 and 300. The BN internal
memory duration, $\tau$, is commensurable with halfdecay time of the excitatory
postsynaptic potentials (EPSP). Thus, in the framework of this extremely simple model,
one could expect that the ability of biological neural network to condense information
should depend on its geometric size, or on the relationships between interneuronal
transmission delays and the EPSP halfdecay time.
\bigskip

{\small\bf Acknowledgments}
\noindent{\small
This work was supported by the Program of basic research of the National Academy of Science of
Ukraine.

Content of this work was partially published in an abstract form in the abstract 
book of the 2nd International Biophysics Congress and Biotechnology at 
GAP \& 21th National Biophysics Congress, (5-9 Oct. 2009)
Diyarbak\i r, Turkey, \verb+http://www.ibc2009.org/+
}

\end{multicols}

\medskip



\begin{thebibliography}{000}


\bibitem{Acharya0}
Acharya, R., Chua, E.C.P., Chua, K.C., Min, L.C., and Tamura, T. (2010), “Analysis and Automatic Identification of Sleep Stages using Higher Order Spectra,” International Journal of Neural Systems, 20:6.

\bibitem{Bullier}
J. Bullier,
``Integrated model of visual processing,''
{\em Brain Res. Rev.,} {\bf 36}, 96--­107 (2001).

\bibitem{Merzenich}
D. V. Buonomano and M. M. Merzenich,
``Cortical plasticity: from synapses to maps,''
{\em Annu. Rev. Neurosci.,} {\bf 21}, 149--186 (1998).

\bibitem{Cariani}
P. Cariani,
``Temporal codes, timing nets, and music perception,''
{\em J. New Music Res.,} {\bf 30}, 107--­135 (2001).

\bibitem{Damasio89}
A. R. Damasio, 
``The brain binds entities and events by multiregional activation from convergence zones,''
{\em Neural Comput.,} {\bf 1}, 123--132 (1989).

\bibitem{Viret}
P. Duchamp-Viret and A. Duchamp,
``Odor processing in the frog olfactory system,''
{\em Prog. Neurobiol.,} {\bf 53}, 561--602 (1997).

\bibitem{Durbin}
R. Durbin and G.  Mitchison,
``A dimension reduction framework for understanding cortical maps,''
{\em Nature,} {\bf 343}, 644--­647 (1990).

\bibitem{Eckhorn}
R. Eckhorn, R. Bauer, W. Jordan, M. Brosch, W. Kruse, M. Munk and H. J. Reitboeck, 
``Coherent oscillations: a mechanism for feature linking in the visual cortex?,''
{\em Biol. Cybern.,} {\bf 60}, 121--130 (1988).

\bibitem{Engel91a}
A. K. Engel, P. K\"onig, A. K. Kreiter, C. M. Gray and W. Singer, 
``Temporal coding by coherent oscillations as a potential solution to the binding problem: physiological evidence,''
In Schuster, H.G., Singer, W. (ed), 
{\em Nonlinear Dynamics and Neuronal Networks,} 3--25. 
VCH Weinheim (1991).

\bibitem{Feldman}
J. Feldman,
``Ecological expected utility and the mythical neural code,''
{\em Cognitive Neurodynamics,} {\bf 4}, 25--35 (2010).

\bibitem{Adeli0}
Ghosh-Dastidar, S. and Adeli, H. (2007), “Improved Spiking Neural Networks for EEG Classification and Epilepsy and Seizure Detection,” Integrated Computer-Aided Engineering, Vol. 14, No. 3,  pp.  187-212.

\bibitem{Adeli}
S. Ghosh-Dastidar and H. Adeli,
``Spiking neural networks,''
{\em Intern. J. Neural Sys.} {\bf 19}, 295--308 (2009).

\bibitem{Adeli2}
S. Ghosh-Dastidar and H. Adeli, “A New Supervised Learning Algorithm for Multiple Spiking Neural Networks with Application in Epilepsy and Seizure Detection,” 
{\em Neural Networks} {\bf 22}, 1419--1431 (2009).

\bibitem{Gorban}
A. Gorban and A. Zinovyev,
``Principal manifolds and graphs in practice: from molecular biology to dynamical systems," {\em Intern. J. Neural Sys.,} {\bf 20}, 219--232 (2010).

\bibitem{self}
H. Haken, ``Brain Dynamics. Synchronization and Activity Patterns in Pulse-Coupled Neural Nets with Delays and Noise,'' Springer, Berlin 2007.

\bibitem{Villa}
J. Iglesias and A. E. P. Villa,
``Emergence of preferred firing sequences in large spiking neural networks during simulated neuronal development,''
{\em Intern. J. Neural Sys.,} {\bf 18}, 267--277 (2008).

\bibitem{Johnston0}
Johnston, S.P., Prasad, G.,Maguire, L. and McGinnity, T.M. (2010), “An FPGA Hardware/software co-design methodology -  towards evolvable spiking networks for robotics application,” {\em Intern. J. Neural Sys.}, 20:6.   

\bibitem{Schwartz}
E. R. Kandel, J. H. Schwartz and T. M. Jessell, 
``Principles of Neural Science,'' (Fourth ed.), New York: McGraw-Hill (2000).

\bibitem{Thorpe1}
H. Kirchner and S. J. Thorpe,
``Ultra-rapid object detection with saccadic eye movements: visual processing speed revisited,''
{\em Vision Res.,}  {\bf 46}, 1762--1776 (2006).

\bibitem{Konig}
P. K\"onig, A. K. Engel and W. Singer,
``Integrator or coincidence detector? The role of the cortical neuron revisited,''
{\em Trends Neurosci.,} {\bf 19}, 130--137 (1996).

\bibitem{KonigKruger}
P. K\"onig and N. Kr\"uger,  
``Symbols as self-emergent entities in an optimization process of feature extraction and predictions,''
{\em Biol. Cybern.,} {\bf 94}(4), 325--334 (2006).

\bibitem{MacKay0}
D. M. MacKay  and W. S. McCulloch, 
``The limiting information capacity of a neuronal link,''
{\em Bull. Math. Biophys.,} {\bf 14}, 127--135, (1952).

\bibitem{MacKay}
D. M. MacKay, 
``Self-organization in the time domain,''
In M. C. Yovitts, G. T. Jacobi and G. D. Goldstein (ed), 
{\em Self-Organizing Systems,}
Spartan Books, Washington, 37--48 (1962).

\bibitem{Nichols0}
Nichols, E., McDaid, L.J., and Siddique, N.H. (2010), “Case Study on Self-organizing Spiking Neural Networks for Robot Navigation,” {\em Intern. J. Neural Sys.}, 20:6.  

\bibitem{Passaglia}
Ch. L. Passaglia and J. B. Troy,
``Information transmission rates of cat retinal ganglion cells,''
{\em J. Neurophysiol.,} {\bf 91}, 1217--­1229 (2004).

\bibitem{Rudolph}
M. Rudolph and A. Destexhe, 
``Tuning neocortical pyramidal neurons between integrators and coincidence detectors, ''
{\em J. Comput. Neurosci.,} {\bf 14}, 239--251 (2003).

\bibitem{Scharf}
M. T. Scharf, N. H. Woo, K. M. Lattal, Z. Young, P. V. Nguyen and T. Abel,
``Protein synthesis is required for the enhancement of long-term potentiation and long-term memory by spaced training,''
{\em J. Neurophysiol.,} {\bf 87} 2770--2777 (2002).

\bibitem{Schliebs0}
Schliebs, S., Kasabv, N., and Defoin-Platel, M. (2010), “On the Probabilistic Optimization of Spiking Neural Networks,” {\em Intern. J. Neural Sys.}, 20:6.  

\bibitem{Shannon}
C. E. Shannon, 
``A mathematical theory of communication,''
{\em Bell System Technical J.,} {\bf 27}, 379--423 and 623--656, July and October, 
(1948).

\bibitem{Soltic0}
Soltic and S. Kasabov, N. (2010), “Knowledge extraction from evolving spiking neural networks with rank order population coding ,” {\em Intern. J. Neural Sys.}, 20:6. 

\bibitem{Strain0}
Strain, T.J., McDaid, L.J., Maguire, L.P., and T.M. McGinnity, T.M. (2010), “An STDP Training Algorithm for a Spiking Neural Network with Dynamic Threshold Neurons,” {\em Intern. J. Neural Sys.}, 20:6.  

\bibitem{Tang}
S. Tang, R. Wolf, S. Xu and M. Heisenberg,
``Visual pattern recognition in Drosophila is invariant for retinal position,''
{\em Science,} {\bf 305}, 1020--1022 (2004).

\bibitem{Thorpe}
S. Thorpe, D. Fize and C. Marlot,
``Speed of processing in the human visual system,''
{\em Nature,} {\bf 381}, 520--522 (1996).

\bibitem{Vid96}
A. K. Vidybida,
``Neuron as time coherence discriminator,''
{\em Biol. Cybern.,} {\bf 74}, 539--544 (1996).

\bibitem{Vid96a}
A. K. Vidybida,  ``Information processing in a pyramidal-type neuron,'' {\it Proc. 
BioNet'96 ­ Biologieorientierte Informatik und pulspropagierende Netze, 3-d Workshop.} Ed.: Heinz G., pp. 96--99. (1996).

\bibitem{Vid98}
A. K. Vidybida, 
``Inhibition as binding controller at the single neuron level,''
{\em BioSystems,} {\bf 48}, 263--267 (1998).

\bibitem{BNF}
A. K. Vidybida, 
``Output stream of binding neuron with instantaneous feedback,''
{\em Eur. Phys. J. B,} {\bf 65}, 577--584 (2008); {\em Eur. Phys. J. B,} {\bf 69}, 313 (2009).

\bibitem{Laurent}
M. Wehr and G. Laurent,
``Odor encoding by temporal sequences of firing in oscillating neural assemblies''
{\em Nature,} {\bf 384}, 162--166 (1996).

\bibitem{Widrow}
B. Widrow,
``Generalization and information storage in networks of adaline `neurons' ''
in: Yovitz,M.C., Jacobi,G.T., Goldstein,G.(ed),
{\em Self-Organizing Systems,} 435--461 (1962).

\bibitem{Wu}
W. Wu and T. Chen,
``Impossibility of asymptotic synchronization for pulse-coupled oscillators with   delayed excitatory coupling,"
             {\em Intern. J. Neural Sys.,} {\bf 19}, 425--435 (2009).



\phantom{00}
\end{thebibliography}
\end{document}